\documentclass[reprint, amsmath,amssymb, aps,]{revtex4-1}

\usepackage{amssymb}
\usepackage[caption=false]{subfig}
\usepackage{amssymb}
\usepackage{amsmath}
\usepackage{graphicx}
\usepackage[caption=false]{subfig}
\usepackage{verbatim}
\usepackage{bm}
\usepackage[normalem]{ulem}
\usepackage[utf8]{inputenc}
\usepackage{caption}

\begin{document}

\title{Slowly rotating topological neutron stars -- universal relations and epicyclic frequencies}

\author{Victor~I.~Danchev$^1$}
\homepage{Electronic address: \href{vidanchev@uni-sofia.bg}{vidanchev@uni-sofia.bg}}
\author{Daniela~D.~Doneva$^{2,3}$}
\homepage{Electronic address: \href{daniela.doneva@uni-tuebingen.de}{daniela.doneva@uni-tuebingen.de}}
\author{Stoytcho~S.~Yazadjiev$^{1,4}$}
\homepage{Electronic address: \href{yazad@phys.uni-sofia.bg}{yazad@phys.uni-sofia.bg}}
\affiliation{$^1$Department of Theoretical Physics, Faculty of Physics, Sofia University, Sofia 1164, Bulgaria}
\affiliation{$^2$Theoretical Astrophysics, Eberhard Karls University of Tübingen, Tübingen 72076, Germany}
\affiliation{$^3$INRNE -- Bulgarian Academy of Sciences, 1784 Sofia, Bulgaria}
\affiliation{$^4$Institute of Mathematics and Informatics, Bulgarian Academy of Sciences, Acad. G. Bonchev Street 8, Sofia 1113, Bulgaria}

\date{\today}

\begin{abstract}
In the modern era of abundant X-ray detections and the increasing momentum of gravitational waves astronomy, tests of General Relativity in strong field regime become increasingly feasible and their importance for probing gravity cannot be understated.
To this end, we study the characteristics of slowly rotating topological neutron stars in the tensor-multi-scalar theories of gravity following the static study of this new type of compact objects by two of the authors.
We explore the moment of inertia and verify that universal relations known from General Relativity hold for this new class of compact objects.
Furthermore, we study the properties of their innermost stable circular orbits (ISCO) and the epicyclic frequencies due to the latter's hinted link to observational quantities such as quasi-periodic X-ray spectrum features.
\end{abstract}

\maketitle


\section{Introduction}\label{introduction}
Some of the most promising extensions of General Relativity (GR) are based on the existence of additional fields which may contribute to Gravity either as additional energy-momentum sources or even through direct coupling to it. While weak-field tests place strong restrictions to such extensions, these fields may have a strong impact on the properties of compact objects such as neutron stars and black holes, which places these objects among the best ``laboratories'' for experimental tests of the strong field regime of gravity \cite{Barack_2019, Berti_2015, Doneva:2017jop}.
The Tensor-Multi-Scalar Theories (TMST) are among the most promising and natural such extensions of General Relativity which are mathematically self-consistent and pass all known experimental and observational constraints \cite{Damour,Horbatsch}.
In this paper we explore some of the most important observational properties of slowly rotating stars in a certain class of TMST.
A new type of neutron stars called topological neutron star was shown to exist in their framework \cite{Yazad_TNS} (see also \cite{Doneva:2020afj} for the case of scalarization in such theories). Following the study of the static structure of this new type of compact objects, we further consider their properties under slow rotation in order to extract additional observational signatures.

One of the most important quantities which can be obtained under rotation is the moment of inertia, which can be computed to first order in the star's angular velocity.
Due to the largely uncertain equation of state (EOS) for extremely dense matter, it is beneficial to construct universal relations which provide links between the parameters of neutron stars independently of the EOS.
We will focus on two types of universal relations connecting the suitably normalized moment of inertia and the stellar compactness. These relations were examined in the GR case in \cite{Lattimer_1,Breu_1}. Later they were generalized to the case of $f(R)$ and scalar-tensor theories with a massive scalar field \cite{Doneva,Kalin_2_Uni,Popchev_2018}.
We follow their methodology and compare the topological neutron star results with the GR case for several EOS and at several theory parameters for two coupling functions in order to discover possible observational traits of the former. 

The modern and future X-ray observatories such as NICER, LOFT and SKA \cite{NICER_1,SKA} can give us invaluable data for neutron stars possessing less compact companions such as white dwarfs or main sequence stars, due to the high likelihood of accretion in such systems. One of the most important parameters that characterize such an accretion is the innermost stable circular orbit (ISCO) since it determines the boundary region where the accreting matter can no longer orbit the compact object under gravity alone. The epicyclic frequencies for a given orbital radius are the characteristic frequencies of oscillation for particles on stable circular orbits undergoing some small perturbations. There are many theories as to the origin of some quasi-periodic oscillations (QPOs) observed in the X-ray light curves of accreting compact objects, but there is no consensus which is the correct one (see \cite{vdKlis} for a comprehensive review). The oscillations themselves are in the order of a few hundred Hz to kHz and most models are based either on orbital and epicyclic motion of matter \cite{Miller,Kalin_2015} or oscillations and instabilities in an accretion disc \cite{Rezzolla_1,deAveller} both of which are expected to occur close to the ISCO radius. Nevertheless, almost all QPOs models are related in some way or another to the ISCO radius and the epicyclic frequencies. Since  ISCO is located close (or on) the surface of neutron stars, the QPOs are a promising opportunity to probe the strong regime of gravity and its potential modifications. The QPOs in scalar-tensor type alternative theories of gravity were considered in \cite{DeDeo:2004kk}--\cite{Kalin_2019}. In the present paper we will address this problem in the context of TMST and topological neutron stars.

We start with the standard static spherical metric, perturbing it to first order in the angular velocity $\Omega$ and solving the resulting system of reduced field equations. The final goal is to extract the moment of inertia for the topological neutron stars and compare two universal (independent of the EOS) relations, known to be valid in GR \cite{Lattimer_1,Breu_1} and some alternative theories of gravity \cite{Doneva,Popchev_2018}. Furthermore, we compute the radius of the innermost stable circular orbit (ISCO), the epicyclic frequencies and their dependence on topological neutron star mass and angular velocity for the stable branches of solutions for one of the coupling functions considered.

Inline with the previous work, we consider a gravitational interaction mediated by the spacetime metric $g_{\mu \nu}$ and $N$ scalar fields $\varphi^a$ with values in a coordinate patch of an N-dimensional Riemannian (target) manifold $\mathcal{E}_N$ with a positive-definite metric $\gamma_{a b}(\varphi)$ defined on it.
The Einstein frame action  of the theory is given by
\begin{eqnarray}\label{action}
S = && \frac{1}{16\pi G_{*}}\int d^4 \sqrt{ - g }[ R - 2 g^{\mu \nu} \gamma_{ a b } \nabla_{\mu} \varphi^{a} \nabla_{\nu} \varphi^{b} - 4 V(\varphi) ] \nonumber\\
&& + S_{\mathrm{matter}}( A^2(\varphi) g_{\mu \nu} , \Psi_{\mathrm{matter}} ),
\end{eqnarray}
where $G_{*}$ is the bare gravitational constant, R and $\nabla_{\mu}$ are the Ricci scalar and the covariant derivative with respect to the Einstein frame metric $g_{\mu \nu}$, and $V(\varphi) \geq 0$ is the scalar fields potential.

The matter fields are coupled only to the physical Jordan frame metric $\tilde{g}_{\mu \nu} = A^2(\varphi) g_{\mu \nu}$ in order for the theory to satisfy the weak equivalence principle.
The conformal factor $A^2(\varphi)$, the target space metric $\gamma_{a b}$ and the scalar fields potential $V(\varphi)$ specify the TMST.

The variation of (\ref{action}) with respect to the Einstein frame metric components and the scalar fields gives the field equations of the theory in the Einstein frame as follows
\begin{eqnarray}\label{field_eqs}
R_{\mu \nu} = &&  8 \pi G_{*} \left( T_{\mu \nu} - \frac{1}{2} T g_{\mu \nu} \right) + \nonumber\\
&& 2 \gamma_{a b}(\varphi)\nabla_{\mu} \psi^{a} \nabla_{\nu} \psi^{b} + 2 V(\varphi) g_{\mu \nu}, \nonumber\\
\nabla_{\mu} \nabla^{\mu} \varphi^{a} = && - \gamma^{a}_{b c}(\varphi) g^{\mu \nu} \nabla_{\mu} \varphi^{b} \nabla_{\nu} \varphi^{c} + \gamma^{a b}( \psi ) \frac{\partial V(\varphi)}{\partial \varphi^b}  \\
&& - 4\pi G_{*} \gamma^{a b}(\varphi) \frac{\partial \ln A(\varphi) }{\partial \varphi^b} T, \nonumber
\end{eqnarray}
where $T_{\mu \nu} $ is the Einstein frame energy-momentum tensor of the matter and $\gamma^{a}_{bc}(\varphi)$ are the Christoffel symbols with respect to the target space metric $\gamma_{a b}$, following closely the notation in \cite{Yazad_TNS}.
The conservation law for the energy-momentum tensor obtained from the contracted Bianchi identities and the field equations is
\begin{eqnarray}\label{EM_conservation}
\nabla_{\mu} T^{\mu}_{\nu} = \frac{\partial \ln A(\varphi) }{\partial \varphi^a} T \nabla_{\nu}\varphi^a ,
\end{eqnarray}
where once again the physical energy-momentum tensor in the Jordan frame $\tilde{T}_{\mu \nu}$ and the Einstein frame one in (\ref{EM_conservation}) are related through the conformal factor as $T_{\mu \nu} = A^2(\varphi) \tilde{T}_{\mu \nu}$.
The matter fields are described by a perfect fluid and by virtue of this relation, the corresponding energy density, pressure and 4-velocity transformations between the two frames are given by $\varepsilon = A^4(\varphi) \tilde{\varepsilon}$, $p = A^4(\varphi)\tilde{p}$ and $u_{\mu} = A^{-1}(\varphi)\tilde{u}_{\mu}$.

\section{Structure equations for the neutron stars and setting the theory parameters}\label{theory}
We perturb the static spherical solution by a term scaling as $O(\Omega)$ in the Einstein frame
\begin{eqnarray}\label{Line_Element}
ds^2 =&&- e^{2 \Gamma}dt^2 + e^{ 2 \Lambda}dr^2 + r^2( d\theta^2 + \sin^2{\theta} d\phi^2 ) \nonumber\\
&& - 2 r^2 \omega \sin^2{\theta} dt d\phi + O(\Omega^2),
\end{eqnarray}
where $\Gamma$, $\Lambda$ and $\omega$ depend on the radial coordinate $r$ only.
Using the standard Hartle procedure \cite{Hartle_1,Hartle_2}, one can obtain that at spatial infinity $\omega$ tends towards
\begin{eqnarray}\label{Omega_Asymptotics}
\omega(r)  \cong - \frac{2 J}{r^3},
\end{eqnarray}
where $J$ is the angular momentum of the star.

Following the previous work on topological neutron stars, we consider the target space for the TMST to be the round three-dimensional sphere $\mathbb{S}^3$
\begin{eqnarray}\label{Target_Line_Element}
\gamma_{ a b } d\varphi^a d\varphi^b = a^2 \left[ d\chi^2 + \sin^2{\chi} ( d \Theta^2 + \sin^2{\Theta} d\Phi^2 ) \right],
\end{eqnarray}
defined by a radius $a > 0$ of $\mathbb{S}^3$ and the standard angular coordinates $\chi$, $\Theta$, $\Phi$ on $\mathbb{S}^3$.
The motivation behind this choice is the fact that $\mathbb{S}^3$ is among the simplest target spaces that allow the existence of spherically symmetric topological neutron stars.
The assumption that the $\chi$ field depends only on the radial coordinate ($\chi = \chi(r)$) and that the $\Theta$ and $\Phi$ fields depend only on the corresponding angular coordinates ($\Theta = \Theta(\theta)$ and $\Phi = \Phi(\phi)$) leads to unique solutions of the latter two, compatible with the spherical symmetry of the structure equations. These solutions are $\Theta = \theta$ and $\Phi = \phi$ as shown in \cite{Yazad_TNS}.

\subsection{Background structure equations}\label{Static_th}
Under these simplifying assumptions, one can readily derive the structure equations for the static configuration. Using the field equations (\ref{field_eqs}) and the conservation law (\ref{EM_conservation}), the reduced field equations take the form:
\begin{widetext}
\begin{eqnarray}\label{Lambda_struc_eq}
&& \frac{2}{r}e^{ - 2\Lambda } \Lambda' + \frac{1}{r^2}\left( 1 - e^{-2\Lambda} \right) = 8\pi G_{*} A^4(\chi) \tilde{\varepsilon} + a^2 \left( e^{-2\Lambda} \chi'^2 + 2 \frac{\sin^2{\chi}}{r^2} \right) + 2 V(\chi), \\ \label{Gamma_struc_eq}
&& \frac{2}{r}e^{ - 2\Lambda } \Gamma' - \frac{1}{r^2}\left( 1 - e^{-2\Lambda} \right) = 8\pi G_{*} A^4(\chi) \tilde{p} + a^2 \left( e^{-2\Lambda} \chi'^2 - 2 \frac{\sin^2{\chi}}{r^2} \right) - 2 V(\chi), \\ \label{Chi_struc_eq}
&& \chi'' + \left( \Gamma' - \Lambda' + \frac{2}{r} \right)\chi' = \left[ 2 \frac{\sin{\chi}\cos{\chi}}{r^2} + \frac{1}{a^2}\frac{\partial V(\chi)}{\partial \chi} + \frac{4 \pi G_{*}}{a^2}A^4( \chi ) \frac{\partial \ln A(\chi)}{\partial \chi}( \tilde{\varepsilon} - 3 \tilde{p} ) \right] e^{2 \Lambda} \\ \label{p_struc_eq}
&& \tilde{p}' = - ( \tilde{\varepsilon} + \tilde{p} )\left[ \Gamma' + \frac{\partial \ln A(\chi)}{\partial \chi} \chi' \right],
\end{eqnarray}
\end{widetext}
where the prime denotes differentiation with respect to the radial coordinate $r$.
Naturally, these equations must be complemented by an appropriate equation of state (EOS) which provides the required dependence between the physical pressure and energy density ($\tilde{p} = \tilde{p}( \tilde{\varepsilon} )$) in order to complete the system.
Throughout the work we use a piecewise polytropic approximation of several realistic nuclear matter equations of state \cite{APR4_poly}. These are the  APR4 and Sly, that fit very well to the current observational constraints, as well as several higher/lower stiffness and maximum mass equations of state (MS1, MPA1, APR2, H4) in order to check the universality of the relations.

Apart from the standard boundary conditions for the metric functions derived from asymptotic flatness $\Gamma( \infty ) = \Lambda( \infty ) = 0$ and regularity at the center $\Lambda(0) = 0$, one further requires such regularity from the scalar field equations. 
This leads to the conditions $\chi(\infty) = k\pi$ and $\chi(0) = n \pi$ with both $k$ and $n$ integer numbers $k, n \in \mathbb{Z}$.
The former condition can be set to zero without loss of generality ($ k = 0 $).
Since $\chi(\infty) = 0$, the extension of the map $\phi : \Sigma \rightarrow \mathbb{ S }^3$ is topologically equivalent to the map $\phi : \mathbb{ S }^3 \rightarrow  \mathbb{ S }^3$, and it can be shown that $n$ is its degree \cite{Yazad_TNS}. Thus the solutions with $n \ne 0$ are topologically nontrivial.

We focus on two different coupling functions characterized by a single dimensionless parameter $\beta$. The first one is that originally used in \cite{Yazad_TNS} given by 
\begin{eqnarray}\label{nmonotonic_coupling}
A(\chi) = e^{ \beta \sin^2{\chi} },
\end{eqnarray}
while the second one is a monotonic function of $\chi$
\begin{eqnarray}\label{monotonic_coupling}
A(\chi) = e^{ \frac{1}{2}\beta \chi^2 }.
\end{eqnarray}
Solutions of the static topological neutron stars are obtained by setting the appropriate central conditions for $\Lambda$, $\tilde{\varepsilon}$ and $\chi$ and performing a shooting method to determine $(d\chi/dr)|_{0}$ and $\Gamma(0)$ from the asymptotics of $\chi$ and $\Gamma$. For the numerical integration of the differential equations we are using the adaptive Dormand-Prince embedded error estimation method \cite{Dormand_Prince}.

As shown in the previous work, solutions exist only for certain theory parameters.
One of the main constraints is the size of the target space $a$ with topological neutron stars emerging only for small values of it.
The radius of the target space throughout this work has been set to $a^2 = 10^{-3}$.
There also exists a restricted range of $\beta$ values at fixed $a$ which depends on the coupling function.
Extending the results obtained previously, several different values of $\beta$ have been used for the function (\ref{nmonotonic_coupling}).
These include $\beta = 0.08$ which was already employed in \cite{Yazad_TNS,Doneva2020} as well as $\beta = 0.04$, $\beta = -0.04$ and $\beta = -0.08$. 
In the case of the coupling function (\ref{monotonic_coupling}), solutions were restricted to an even smaller interval of values for $\beta$ and there is strong evidence that no stable configurations exist for $\beta > 0$ in that case.

\begin{figure}[t]
\centering
\includegraphics[scale=0.55]{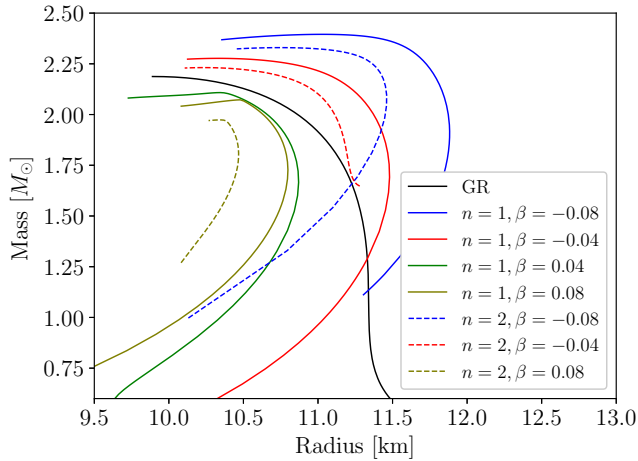}
\caption{\label{MR_Relation_1} Mass-radius relations for the APR4 EOS and different values of $\beta$ for the coupling function (\ref{nmonotonic_coupling}). $n=1$ solutions are shown with solid lines while $n=2$ solutions are indicated with dashed lines of the same color. }
\end{figure}

\begin{figure}[t]
	\centering
	\includegraphics[scale=0.55]{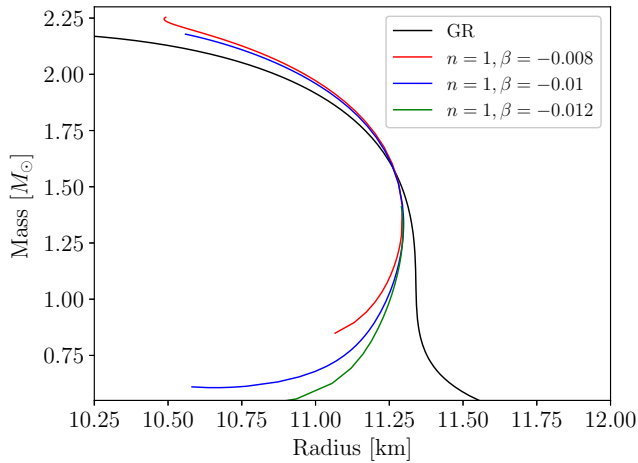}
	\caption{\label{MR_Relation_2} Mass-radius relations for the APR4 EOS and different values of $\beta$ for the coupling function (\ref{monotonic_coupling}). Only $n=1$ stable solutions exist for this coupling function.}
\end{figure}

The mass-radius relations for topological neutron stars is shown in Fig. \ref{MR_Relation_1} for the non-monotonic coupling function (\ref{nmonotonic_coupling}). In this case stable solutions exist both for $n=1$ and $n=2$ topological charges (excluding the case $\beta = 0.04$ where the $n=2$ solutions appear after the branch's maximum mass).
Fig. \ref{MR_Relation_2} on the other hand presents the  mass-radius relations for the monotonic coupling function (\ref{monotonic_coupling}) where stable solutions exist only for $n=1$ topological charge and in a much smaller range of $\beta$ (around $\beta = -0.01$). Since we are working in slow rotation approximation keeping only linear terms with respect to the angular velocity $\Omega$, the mass and the radius remain unchanged with respect to the nonrotating case. The results depicted in these two figures are for the stable branch of topological neutron star solutions. It was shown in \cite{Yazad_TNS} that other branches of solutions exist as well for a fixed coupling function and values of $n$ and $\beta$, but all of them are  unstable against radial perturbations \cite{Doneva2020} and we will not comment them further.

It is easy to notice that in general the $n=1$ branches are more massive than their $n=2$ counterparts (when the latter exist).
Considering Fig.\ref{MR_Relation_1}, the overall effect of negative $\beta$ values is to increase the masses of the solutions with the same radius (making them more compact) while the effect of positive $\beta$ values is the opposite. The actual shape of the mass-radius relation, however, is also heavily influenced by the value of the parameter and these effects are not easily predictable or monotonic in nature.

In most cases the ``end'' \ of a branch (i.e. the highest pressure for which solutions were found) has been chosen arbitrarily after the corresponding sequence's maximum mass is reached. However, in some cases the branch of solutions not only appears at some minimal central pressure but also disappears at another maximum central pressure. This maximum pressure where solutions disappear might even occur before the maximum of the mass is reached. Such is the case for the solutions obtained for the monotonic coupling function (\ref{monotonic_coupling}) which are found in a much more restricted range for $\beta$ and central pressures.
No stable or physically meaningful solutions were found for the $n=2$ topological charge in this case and each of the three stable branches shown on Fig. \ref{MR_Relation_2} disappears before it reaches the maximum of the mass.

\subsection{Structure equations to first order in $\Omega$}\label{Omega_th}
The only non-trivial equation to first order in $\Omega$ is that for the Ricci tensor components $R_{03}=R_{30}$, which after taking into account the static equations (\ref{Lambda_struc_eq}-\ref{p_struc_eq}) as equalities, simplifies to
\begin{eqnarray}\label{Omega_struc_eq}
&& \frac{1}{r^4}\frac{d}{dr}\left( r^4 e^{ - \Gamma - \Lambda }  \bar{\omega}' \right)  + \frac{4}{r}\frac{d}{dr} (e^{ - \Gamma - \Lambda } ) \bar{\omega} \nonumber \\ 
&& + 4 a^2 e^{ - \Gamma - \Lambda }   \bar{\omega} \chi'^2 + \frac{4 a^2 \omega}{r^2} \sin^2{\chi} e^{ \Lambda - \Gamma} = 0,
\end{eqnarray}
where we have defined $ \bar{\omega} = \Omega - \omega $.
The central values of the functions are $\bar{\omega}'(0) = 0$ and $\bar{\omega}(0) = \bar{\omega}_c$ with the latter determined through a shooting method in order to obtain the desired angular velocity of the star $\Omega$.
Note that $\Omega$ is actually the same in both Einstein and Jordan frames.

An important characteristic for compact objects is their moment of inertia. 
Having integrated (\ref{Omega_struc_eq}), one can use its asymptotic form to extract the star's angular momentum $J = I \Omega$ and thus obtain the moment of inertia along the axis of rotation. Alternatively, one can integrate the two sides of (\ref{Omega_struc_eq}) leading to a conserved  quantity proportional to the angular momentum. After a division by $\Omega$, the asymptotic and integral definitions are given by
\begin{eqnarray}\label{MOI_eq}
I_{\mathrm{asympt}} = && \lim_{r \rightarrow \infty} - \frac{1}{6}r^4 e^{ - \Gamma - \Lambda } \left( \frac{d \tilde{\omega}}{d r} \right), \\
I_{\mathrm{integral}} = && \frac{8 \pi G_{*}}{3} \int_{0}^{r_S}  A^4(\chi) (\tilde{p} + \tilde{\varepsilon} ) e^{\Lambda - \Gamma} r^4( 1 - \tilde{\omega} ) dr \nonumber\\
&& - \frac{2 a^2}{3} \int_{0}^{\infty} r^2 \tilde{\omega} e^{\Lambda - \Gamma} \sin^2{\chi} dr
\end{eqnarray}
where $\tilde{\omega} = \omega/\Omega = 1 - \bar{\omega}/\Omega $ is the reduced angular velocity.
Both definitions are implemented as part of the numerical integration to further verify the fidelity of the results and the obtained values differ from each-other with relative error consistently on the order of $10^{-11}$ for the various branches.
Section \ref{Uni_relations_res} outlines our results for the moment of inertia and its normalization conditions which lead to universal relations.

\subsection{ISCO, orbital and epicyclic frequencies}\label{ISCO}

For a general stationary axially-symmetric metric with $g_{\mu \nu} = g_{\mu \nu}(r , \theta)$ and a line element of the form
\begin{eqnarray}\label{general_metric}
ds^2\! = \! g_{tt} dt^2 \!  + \! g_{rr} dr^2 \! +\!  g_{\theta \theta} d\theta^2 \! +\!  2 g_{t \phi} dt d\theta \!  +\!  g_{\phi \phi} d\phi^2,
\end{eqnarray}
the ISCO and epicyclic frequencies can be found by analysing the orbital motion for massive test particles.
It is easy to show that there are two constants of motion generated by the timelike and axial killing vectors $ E = - u_t $ and $ L = u_{\phi} $.
Raising these we obtain for the contravariant components of a particle's 4-velocity
\begin{eqnarray}
\frac{d t}{d \tau } = \frac{E g_{\phi \phi} + L g_{t \phi} }{g^2_{t \phi} - g_{t t} g_{\phi \phi} }, \\
\frac{d \phi}{d \tau} = - \frac{E g_{t \phi} + L g_{t t} }{g^2_{t \phi} - g_{t t} g_{\phi \phi}}. 
\end{eqnarray}
The 4-velocity's normalization condition for timelike observers $ g_{\mu \nu}u^{\mu} u^{\nu} = -1 $, can be rewritten in the form 
\begin{eqnarray}\label{Eq_Motion}
g_{r r} \dot{r}^2 + g_{\theta \theta} \dot{\theta}^2 + E^2 U( r , \theta ) = -1,
\end{eqnarray}
where we have defined the potential
\begin{eqnarray}\label{U_pot}
U( r , \theta ) = \frac{g_{\phi \phi} + 2 l g_{ t \phi} + l^2 g_{t t} }{g^2_{t \phi} - g_{t t} g_{\phi \phi}},
\end{eqnarray}
using the proper orbital angular momentum $l \equiv L/E$.
In the equatorial plane ($\theta = \pi/2$) this result is further reduced to an effective 1D problem $ \dot{r}^2 = V(r) $ with effective potential
\begin{eqnarray}\label{V_pot}
V(r) = g_{rr}^{-1} \left[ -1 - E^2 U( r , \theta =\frac{\pi}{2} ) \right].
\end{eqnarray}
For some fixed $E$ and $L$, the stable circular orbit at some coordinate radius $r_0$ is given by the conditions $V(r_0) = V'(r_0) = 0$ and $V''(r_0) > 0$, while the radius of the ISCO is given by the marginal stability condition $V''(r_0) = 0$.
The orbital angular velocity of massive particles in the geometry can be found from the geodesic equations written in their Lagrange-Euler form
\begin{eqnarray}\label{Geodesic_1}
\frac{d}{d\tau}\left( g_{\mu \nu} \frac{d x^{\nu} }{d \tau} \right) = \frac{1}{2} \partial_{\mu} g_{\nu \sigma} \frac{d x^{\nu}}{d\tau}\frac{d x^{\sigma}}{d\tau}.
\end{eqnarray}
The radial component of these equations ($\mu = 1$) reads
\begin{eqnarray}\label{Geodesic_r}
\partial_r g_{t t} \left( \frac{d t}{d \tau}\right)^2 + 2 \partial_r g_{t \phi} \frac{d t}{d \tau} \frac{d \phi}{d \tau} + \partial_r g_{\phi \phi} \left( \frac{d \phi}{d\tau}\right)^2 = 0,
\end{eqnarray}
which is transformed into a quadratic algebraic equation for the orbital angular velocity $\Omega_p = d\phi/dr = u^{\phi}/u^{t}$.
The two solutions are easily found to be
\begin{eqnarray}\label{Orbital_Omega}
\Omega_p = \frac{d\phi}{d t} = \frac{ - \partial_r g_{t \phi} \pm \sqrt{ (\partial_r g_{t \phi} )^2 - \partial_r g_{t t} \partial_r g_{\phi \phi} } }{\partial_r g_{\phi \phi}},
\end{eqnarray}
corresponding to prograde and retrograde orbits with respect to the star's rotation. 
The follow-up computations are considered for the prograde case as the retrogade one's angular velocity is always lower and stability is lost further out from the star.

For a particle on a stable circular orbit, small radial or angular perturbations will cause periodic oscillations about the potential's minimum (with respect to $r$ or $\theta$ respectively).
These frequencies are known as the radial and vertical epicyclic frequencies and can be found by investigating time-dependent perturbations of a stable equatorial circular orbit in the form
\begin{eqnarray}\label{Perturb_Anzatz}
r(t) = r_0 + \delta r(t), \ \ \theta(t) = \frac{\pi}{2} + \delta \theta(t).
\end{eqnarray}
Inserting (\ref{Perturb_Anzatz}) into the equation of motion (\ref{Eq_Motion}) and assuming $\delta r \propto  e^{i \omega_r t} $, $\delta \theta \propto e^{ i \omega_{\theta} t}$ yields
\begin{eqnarray}\label{omega_r}
\omega^2_r = \frac{( g_{t t} + \Omega_p g_{t \phi} )^2}{2 g_{r r}}\partial^2_r U\left( r_0 , \frac{\pi}{2} \right), \\ \label{omega_th}
\omega^2_{\theta} = \frac{( g_{t t} + \Omega_p g_{t \phi} )^2}{2 g_{\theta \theta}}\partial^2_{\theta} U\left( r_0 , \frac{\pi}{2} \right),
\end{eqnarray}
for the radial and vertical angular epicyclic frequencies ($\omega_i = 2 \pi \nu_i$).
Given the interpretation of these frequencies and the proportionality between $\omega_r$ and $\partial^2_r U(r_0 , \pi/2)$, it is clear that the radial epicyclic frequency must be zero at the ISCO radius, real for $r > r_{\mathrm{ISCO}}$ and imaginary for $r < r_{\mathrm{ISCO}}$.
Furthermore, the vertical epicyclic frequency is equal to the orbital one $\omega_{\theta} = \Omega_{p}$ for the static case ($\Omega = 0$).

Of course, realistic neutron stars possess magnetic fields which can be far from negligible on the dynamics of charged particles and recent studies show that those can be more complicated than originally anticipated \cite{NICER_2}.
Nevertheless, since plasma is electrically neutral on the scale of several Debye radii, each ``clump'' of plasma would be held together by electrostatic stresses between the constituting particles so even though individual charged particles' trajectories may differ due to the magnetic field, the overall plasma motion is well described by the considered dynamics.

The same general methodology is also followed for TMST with one major difference -- all derivatives and quantities in the latter case must be computed using the physical (Jordan) frame metric components $\tilde{g}_{\mu \nu}$.
This leads to a noteworthy behavior in the orbital angular velocity $\Omega_p$ as given by (\ref{Orbital_Omega}) which can turn imaginary (i.e. no stable circular orbits exist) in regions outside the star, before the appearance of a true ISCO.
This can be traced to the negative term under the square root of eq. (\ref{Orbital_Omega}).
In GR the derivative $\partial_r g_{tt}$ is always negative outside of the star (the metric function $g_{tt}$ is monotonically decreasing).
However, the sign of the corresponding quantity in the generalized theory $\partial_r (A^2( \chi) g_{tt} ) = \partial_r \tilde{g}_{tt} $ is now determined not only by $g_{tt}$ but also by the coupling function $A( \chi )$.
It turns out that in the case of the non-monotonic coupling function (\ref{nmonotonic_coupling}) this leads to regions outside the ISCO where stable circular orbits do not exist and where the accretion disc may be split.
This effect presents considerable interest in terms of potential observational traits but requires a more detailed study. For that reason it will be explored in a separate work.
In section \ref{ISCO_res} therefore, we outline our results for the radius of ISCO, the orbital frequency at ISCO as well as the maximum value of the radial epicyclic frequency for the $n=1$ stable topological neutron stars of the monotonic coupling function (\ref{monotonic_coupling}) at three different $\beta$ values and three different rotation rates, comparing them to the results for GR.

\section{Moment of Inertia and universal relations results}\label{Uni_relations_res}

In this section we present the results obtained for the moment of inertia as well as our study of two relations known to be universal (independent of EOS).
These are relations between the dimensionless compactness $M/R$ and two normalizations of the moment of inertia $\tilde{I}=I/MR^2$ and $\bar{I}=I/M^3$.
The functional forms were fitted with a polynomial though least squares. Several EOS were used and the error of those fits was evaluated in order to verify their validity.
As discussed, we focus only on the astrophysically relevant stable branch of topological neutron stars without commenting on the other unstable branches discovered in \cite{Yazad_TNS}. A comparison is made between GR and the branches with $n=1$ and $n=2$ topological charges (where present) using APR4, Sly, MS1, MPA1, APR2 and H4 equations of state for several different theory parameters.

Before considering these results for the different EOS, the following Figs. \ref{MI_Relation_1} and \ref{MI_Relation_2} show the moment of inertia as a function of the neutron star masses. These results are obtained for APR4 EOS and several combinations of $n$ and $\beta$ in order to provide some intuition on the effect of the coupling function and the $\beta$ value. 
As a matter of fact these are the same branches of solutions depicted in Fig. \ref{MR_Relation_1} and Fig. \ref{MR_Relation_2}.
\begin{figure}[h!]
\centering
\includegraphics[scale=0.55]{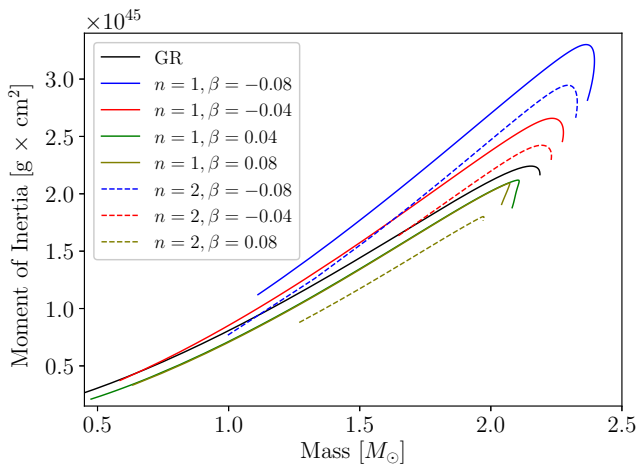}
\caption{\label{MI_Relation_1} Moment of inertia-mass relations for the APR4 EOS and different values of $\beta$ for the coupling function (\ref{nmonotonic_coupling}). $n=1$ solutions are shown with solid lines while $n=2$ solutions are indicated with dashed lines of the same color. These are the same branches of solutions depicted in Fig. \ref{MR_Relation_1}.}
\end{figure}

\begin{figure}[h!]
\centering
\includegraphics[scale=0.55]{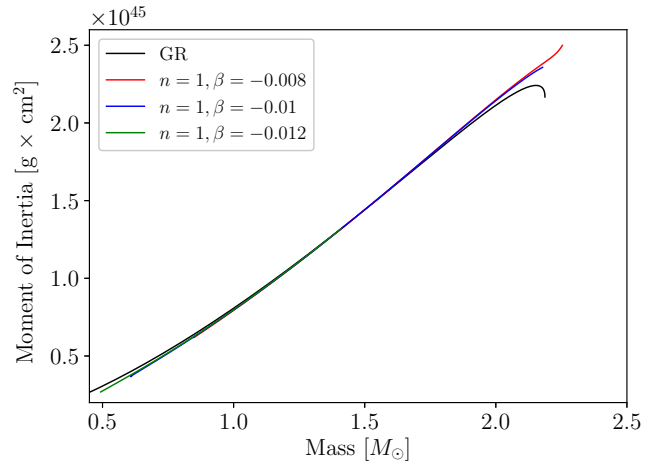}
\caption{\label{MI_Relation_2} Moment of inertia-mass relations for the APR4 EOS and different values of $\beta$ for the coupling function (\ref{monotonic_coupling}). Only $n=1$ stable solutions exist for this coupling function.  These are the same branches of solutions depicted in Fig. \ref{MR_Relation_2}.}
\end{figure}
The behavior of the moment of inertia evidently follows what was already outlined for the mass--radius relation. The higher topological charge leads to lower masses and thus lower moment of inertia. Negative $\beta$ increases the overall compactness and positive $\beta$ acts in the opposite way (for the non-monotonic coupling function (\ref{nmonotonic_coupling})).  A complete lack of maximum of $I(M)$ is observed in the case of the monotonic coupling function. 
It is important to note, that up to first order of the angular velocity, the moment of inertia for a star is independent of $\Omega$ as no deformations are taken into consideration.
We have confirmed that the value for $I$ is indeed numerically independent of the different $\Omega$ values set in our integrator.

Let us now turn to the universal relations. While in the case of mass--radius relations the plots are qualitatively different in shape from those of the GR case (see Figs. \ref{MR_Relation_1} and \ref{MR_Relation_2}), the normalized moment of inertia for all theory parameters follows a very similar dependency. The functional shape is very similar with an overall translation depending on the values of the free parameters (with the exception of $\beta = 0.08, \ n=1$ for the non-monotonic function). 

The fourth order polynomial with zero second and third order terms is now a standard fit for the normalized moment of inertia $\tilde{I}=I/MR^2$ as a function of the compactness $M/R$ \cite{Lattimer_1,Breu_1}
\begin{eqnarray}\label{Uni_Fit1}
\tilde{I}_{\mathrm{fit}} = \tilde{a}_0 + \tilde{a}_1 \frac{M}{R} + \tilde{a}_2 \left( \frac{M}{R}\right)^4
\end{eqnarray}

\begin{figure}[]
\centering
\includegraphics[scale=0.55]{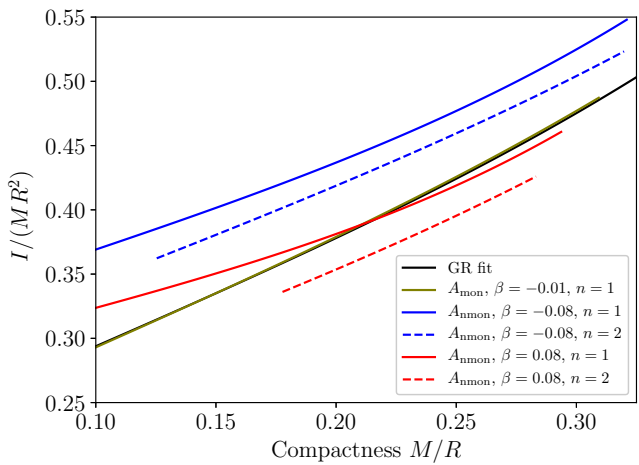}
\caption{\label{Uni_Rel_1_Fits} Comparison between the $\tilde{I}$ fits obtained through (\ref{Uni_Fit1}) for different theory parameters considered. The olive curve is that obtained for the monotonic coupling function (\ref{monotonic_coupling}) while the remaining four curves show the qualitative difference in results for positive and negative beta in the case of non-monotonic coupling (\ref{nmonotonic_coupling}). }
\end{figure}

We will first give plots of the universal relations polynomial fits for different combinations of the free parameters in order to gain intuition on the possible deviations from GR and only afterwards we will comment on the deviation from EOS university. Let us just point out, that this universality is more or less preserved for topological neutron stars as well. 
Fig. \ref{Uni_Rel_1_Fits} compares the fits $\tilde{I}_{\mathrm{fit}}$ for $\beta = -0.01$ of the monotonic coupling function (\ref{monotonic_coupling}), GR and the non-monotonic coupling function (\ref{nmonotonic_coupling}) at values $\beta = -0.08$ and $\beta = 0.08$.
These fits are obtained using an equal number of points for each of the 6 EOS (APR4, SLy, MS1, MPA1, APR2 and H4) and fitting the functional form (\ref{Uni_Fit1}) through least squares. In Table \ref{tbl:Coeff} the numerical values of the fitting coefficients, as well as the fitting error, for some representative combinations of the parameters are displayed.

It is evident both from the table and the plot that there is very little difference between the polynomial fits for GR and the TMST for the monotonic coupling function $A_{\mathrm{mon}}(\chi)$ given by (\ref{monotonic_coupling}). This is not the case, though, for the non-monotonic function  $A_{\mathrm{nmon}}(\chi)$ given by eq. \eqref{nmonotonic_coupling} where significant differences are observed. In this case the normalized moment of inertia is significantly higher or lower as compared to GR both for $n=1$ and $n=2$. For example in the $\beta = -0.08$ case the polynomial fit is consistently around 20 \% higher as compared to GR. 
The $n=1$ fit for the positive $\beta=0.08$ has a slightly different shape from GR particularly at lower compactness but would be hard to discern within experiments available today. 
The $n=2$ solutions of the same value $\beta = 0.08$, however, lead to values of $\tilde{I}$ consistently lower by about 7\%.

The second alternative normalization of the moment of inertia we consider, namely $\bar{I}=I/M^3$, is fitted with a polynomial function of the form \cite{Breu_1}
\begin{eqnarray}\label{Uni_Fit2}
\bar{I}_{\mathrm{fit}} \! = \! \bar{a}_1 \left(\frac{M}{R} \right)^{-1} \! +\! \bar{a}_2 \left(\frac{M}{R} \right)^{-2} \! +\! \bar{a}_3 \left(\frac{M}{R} \right)^{-3}\! + \! \bar{a}_4 \left(\frac{M}{R} \right)^{-4}.
\end{eqnarray}
 The numerical values of the fitting coefficients, as well as the fitting error, can be found once again in Table \ref{tbl:Coeff} for certain representative cases.
 
\begin{figure}[]
\centering
\includegraphics[scale=0.55]{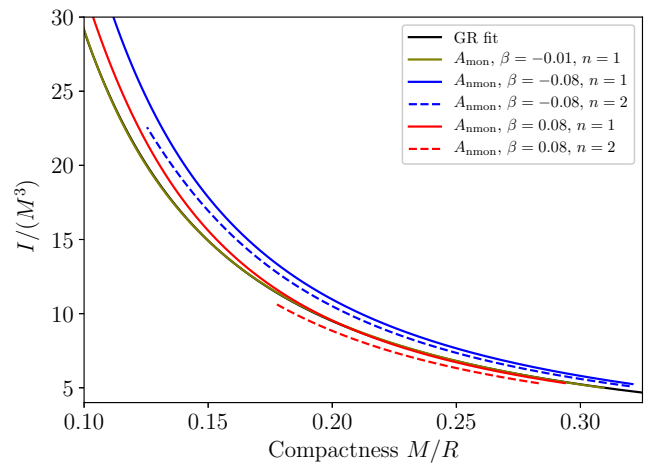}
\caption{\label{Uni_Rel_2_Fits} Comparison between the $\bar{I}$ fits obtained through (\ref{Uni_Fit2}) for different theory parameters considered. The olive curve is that obtained for the monotonic coupling function (\ref{monotonic_coupling}) while the remaining four curves show the qualitative difference in results for positive and negative beta in the case of non-monotonic coupling (\ref{nmonotonic_coupling}).}
\end{figure}

Fig. \ref{Uni_Rel_2_Fits} compares the $\bar{I}_{\mathrm{fit}}$ obtained using the above formula for the same values of the parameters and equations of state as in Fig. \ref{Uni_Rel_1_Fits}.
Once again, the curve for the coupling function (\ref{monotonic_coupling}) is virtually indistinguishable from the GR one and the most significant deviation is observed for the non-monotonic coupling (\ref{nmonotonic_coupling}) with $n=1$ topological charge at $\beta = -0.08$  (about 20 \% higher) and with $n=2$ topological charge at $\beta = 0.08$ (about 7 \% lower).

\begin{figure}[]
\centering
\includegraphics[scale=0.57]{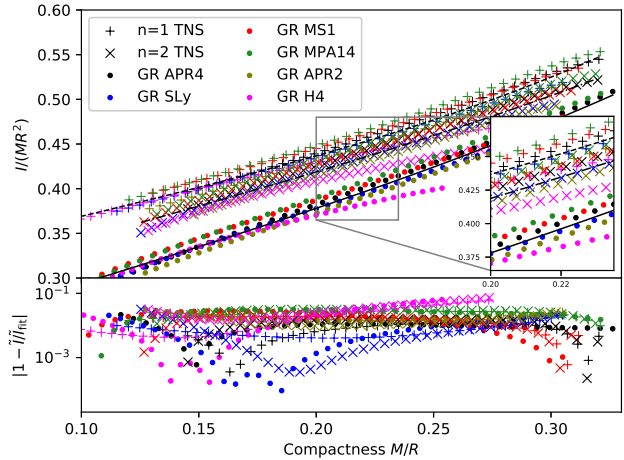}
 \caption{\label{Uni_Rel_1_Neg_Beta} The normalized moment of inertia $\tilde{I}$ as a function of the compactness for  GR and topological neutron stars with a coupling function (\ref{monotonic_coupling}), where we have chosen $\beta = -0.08$. Dots show results for GR while pluses and crosses show results for the topological neutron stars of $n=1$ and $n=2$ respectively, with the same color scheme for the EOS as in GR. Solid, dashed and dash-dotted lines show the corresponding fits following eq. (\ref{Uni_Fit1}).}
\end{figure}

\begin{figure}[h!]
\centering
\includegraphics[scale=0.57]{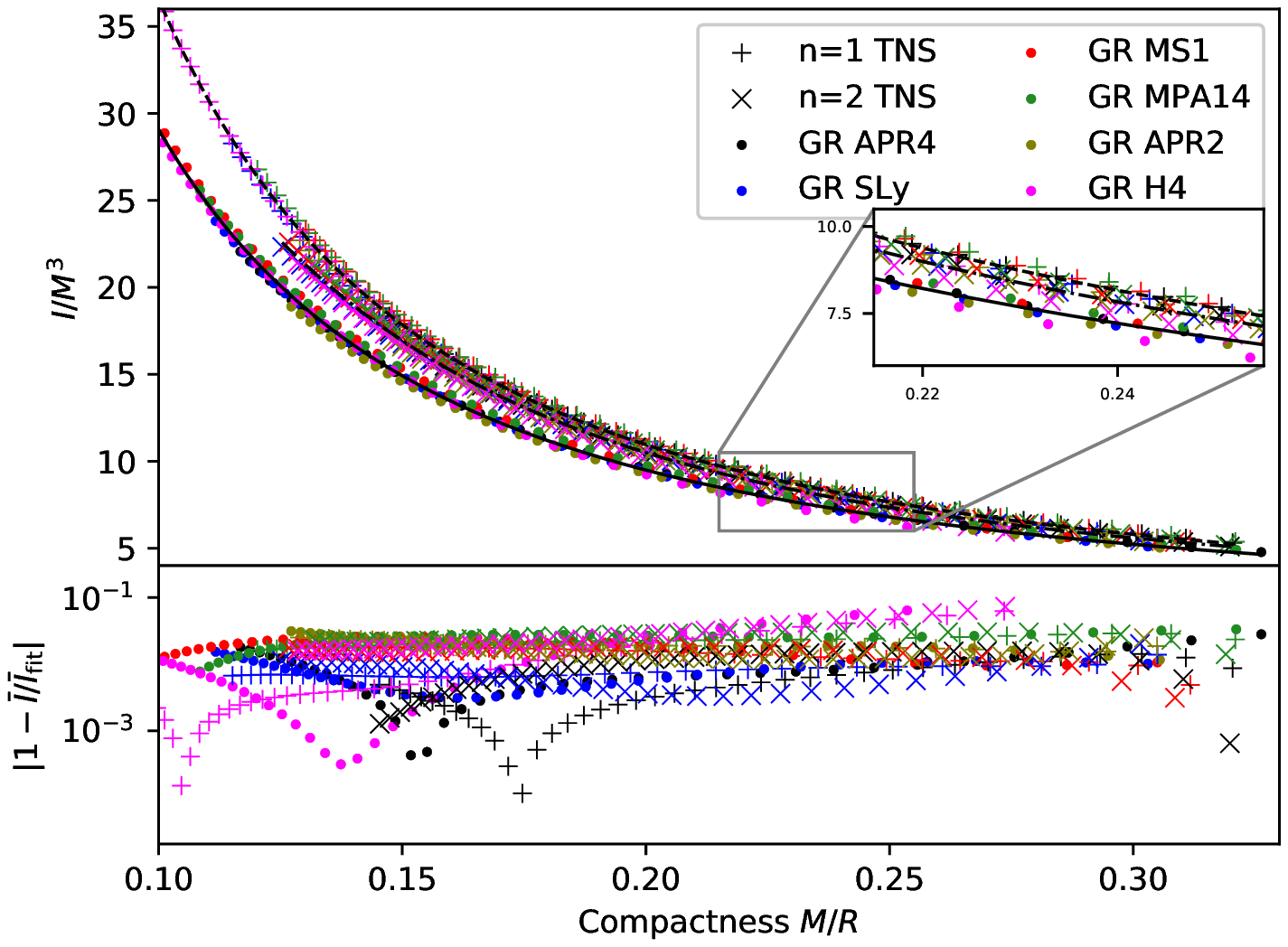}
\caption{\label{Uni_Rel_2_Neg_Beta} The normalized moment of inertia $\bar{I}$ as a function of the compactnessfor  GR and topological neutron stars with a coupling function  (\ref{monotonic_coupling}), where we have chosen $\beta = -0.08$. Dots show results for GR while pluses and crosses show results the Topological Neutron Stars of $n=1$ and $n=2$ respectively with the same color scheme for the EOS. Solid, dashed and dash-dotted lines show the corresponding fits following (\ref{Uni_Fit2}).}
\end{figure}

The following two Figs. \ref{Uni_Rel_1_Neg_Beta} and \ref{Uni_Rel_2_Neg_Beta} show the actual data points for each of the six EOS considered as well as the fits obtained from them.
Only the value of $\beta = -0.08$ of the non-monotonic coupling function (\ref{nmonotonic_coupling}) has been shown for the purpose of better visualization. 
This value of $\beta$ has been chosen since it presents the highest difference as compared to GR. The $n=1$ and $n=2$ solutions are depicted with pluses and crosses and follow the colors of the GR solutions for the different EOS.
The highest relative error of the fits is displayed in the bottom of the figure and it is no greater than 6\% (highest for the H4 EOS).
Based on the relative error of the fits, we can conclude that the EOS universality of  $\tilde{I}$ as a function of the stellar compactness is fulfilled in TMST at least as well as in GR, and the differences between GR and TMST fits are significant at least for some of the $\beta$ and $n$ values.

Similarly, as seen on the following Fig. \ref{Uni_Rel_2_Neg_Beta}, the error between the actual results and the fit for the second normalization $\bar{I}$ does not exceed 6\% with the highest maximum of the error once again visible for H4 EOS which confirms the universality of $\bar{I}$ as well for $\beta = -0.08$.
The fits for the remaining $\beta$ values and the monotonic coupling function have a relative error of the same order, confirming that the topological neutron stars of both charges $n=1$ and $n=2$ indeed satisfy the proposed universal relations and in some cases the predicted difference in the obtained fit is significant as compared to GR.

\begin{widetext}
\begin{center}
 \begin{tabular}{|| c | c | c | c || c || c | c | c | c || c ||} 
 \hline
 Theory parameters & $\tilde{a}_0$ & $\tilde{a}_1$  & $\tilde{a}_2$ & $\tilde{I}_{\chi^2}$ & $\bar{a}_1$ & $\bar{a}_2$  & $\bar{a}_3$ & $\bar{a}_4$ & $\bar{I}_{\chi^2}$ \\ [0.5ex] 
 \hline \hline
 General Relativity & $0.213$ & $0.802$ & $2.610$ & $ 5.8 \times 10^{-5} $  & $0.972$ & $0.161$ & $0.655 \times 10^{-2}$ & $ - 0.033 \times 10^{-2} $ & $ 6.3 \times 10^{-2} $ \\
 \hline
 $A_{\mathrm{mon}}$, $\beta = -0.01$, $n=1$ & $0.211$ & $0.823$ & $2.429$ & $ 5.7 \times 10^{-5} $  & $0.804$ & $0.250$ & $- 0.817 \times 10^{-2}$ & $0.042 \times 10^{-2} $  & $ 6.9 \times 10^{-2} $ \\ 
 \hline
 $A_{\mathrm{nmon}}$, $\beta = -0.08$, $n=1$ & $0.307$ & $0.615$ & $4.110$ & $ 7.0 \times 10^{-5} $ & $1.045$ & $0.151$ & $2.015\times 10^{-2}$ & $-0.092\times 10^{-2}$ & $ 4.9 \times 10^{-2} $ \\ 
 \hline
 $A_{\mathrm{nmon}}$, $\beta = -0.08$, $n=2$ & $0.272$ & $0.717$ & $2.116$ & $ 9.3 \times 10^{-5} $ & $1.325$ & $-0.057$ & $6.193\times 10^{-2}$ & $-0.391\times 10^{-2}$ & $ 5.3 \times 10^{-2} $\\ 
 \hline
 $A_{\mathrm{nmon}}$, $\beta = 0.08$, $n=1$ & $0.274$ & $0.488$ & $5.797$ & $ 5.5 \times 10^{-5} $ & $1.299$ & $-0.056$ & $4.649\times 10^{-2}$ & $-0.221\times 10^{-2}$ & $ 3.7 \times 10^{-2} $\\ 
 \hline
 $A_{\mathrm{nmon}}$, $\beta = 0.08$, $n=2$ & $0.208$ & $0.706$ & $2.818$ & $ 4.8 \times 10^{-5} $ & $1.320$ & $-0.107$ & $5.841\times 10^{-2}$ & $-0.381\times 10^{-2}$ & $ 1.9 \times 10^{-2} $\\ 
 \hline
\end{tabular}
\captionof{table}{\label{tbl:Coeff} The  table summarizes all the $\{ \tilde{a}_j \}$ and $\{ \bar{a}_j \}$ \textbf{coefficients} obtained from the fits of the data with polynomials of the form  (\ref{Uni_Fit1}) and (\ref{Uni_Fit2}). We have chosen some representative combinations of parameters for both coupling functions, where $A_{\mathrm{mon}}$ denotes the monotonic coupling \eqref{monotonic_coupling} and $A_{\mathrm{nmon}}$ -- the non-monotonic coupling \eqref{nmonotonic_coupling}.
Each set of values was obtained by taking an equal number of points for each EOS (in order to obtain the same weight) for mass range starting at $ 1 \ M_{\odot}$ and ending at the maximum mass for the corresponding sequence. }
\end{center}
\end{widetext}

It is evident from the values of $\tilde{I}_{\chi^2}$ and $\bar{I}_{\chi^2}$ displayed in Table \ref{tbl:Coeff} that the topological neutron stars exhibit a good degree of EOS universality with respect to the two normalizations of the moment of inertia $\tilde{I} = I/MR^2$ and $\bar{I} = I/M^3$ and therefore, these relations are well described by the standard fits (\ref{Uni_Fit1}) and (\ref{Uni_Fit2}).  
The obtained fits for the monotonic coupling function (\ref{monotonic_coupling}) are practically indistinguishable from GR.
The non-monotonic coupling function (\ref{nmonotonic_coupling}), however, leads to quantitatively significant differences between GR, the $n=1$ and the $n=2$ topological neutron stars independently of the EOS.

\section{ISCO, orbital and epicyclic frequencies results}\label{ISCO_res}
In this section we outline our results for the ISCO as well as the orbital and epicyclic frequencies.
The results are obtained for static as well as $f = 80 $ Hz and $ f = 160 $ Hz rotating neutron stars.
Both of these values for the frequency place the stars at the slowly rotating regime. Unlike the moment of inertia, which is not influenced to the first order in $\Omega$, the innermost stable circular orbit (ISCO) as well as the orbital $\nu_p$ and epicyclic frequencies $\nu_r $, $\nu_{\theta}$ are modified based on the star's rotation.
Since ISCO is the limiting radius where particles can orbit stably, it has important applications to determining the inner edge of accretion discs around compact objects. We should note that all the results presented in this section are the corresponding physical Jordan frame quantities.

As discussed in section \ref{ISCO}, the ISCO computation in the case of a non-monotonic coupling function such as (\ref{nmonotonic_coupling}) is not trivial and can lead to a splitting of the accretion disc outside of the star.
For the present section, therefore, we have considered only the monotonic coupling function (\ref{monotonic_coupling}) while the more complicated case of non-monotonic $A(\chi)$ requires much more profound investigations and will be discussed in a future publication. 

\begin{figure}[h!]
\centering
\includegraphics[scale=0.55]{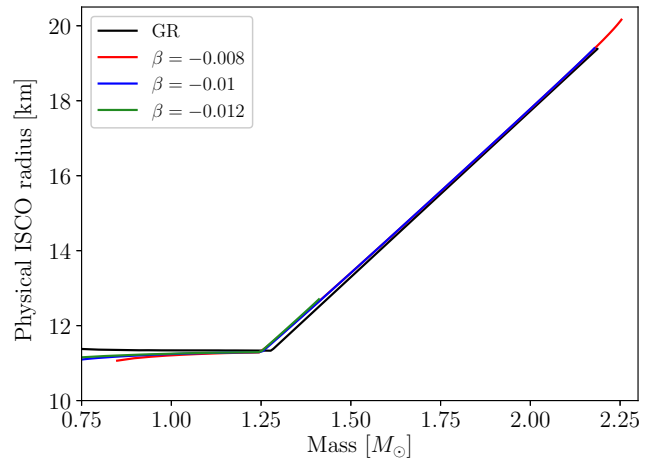}
\caption{\label{ISCO_Static_Plot} ISCO radius (physical) as a function of the gravitational mass for GR and TMST with the monotonic coupling function (\ref{monotonic_coupling}), for $n=1$, different $\beta$ values and $f = 0$ Hz.}
\end{figure}

Fig. \ref{ISCO_Static_Plot} displays the static ISCO radius (in the physical Jordan frame) for the stable branch of topological neutron stars with $n=1$ at different $\beta$ parameters for the APR4 EOS and compares them to GR. Whenever the ISCO radius is less than the stars' radius, the latter is used (thus the visible discontinuity of the derivatives around $1.25 \ M_{\odot}$).
Stable solution for $n=2$ do not exist for this coupling function and as previously discussed, the maximum mass that the neutron stars with $n=1$ and $\beta = -0.012$ can reach, is much lower compared to those for the additional $\beta$ values.
It is obvious that all results follow closely GR and deviations of the ISCO for any physical radii between the considered set of parameters and GR are marginal.

\begin{figure}[h!]
	\centering
	\includegraphics[scale=0.55]{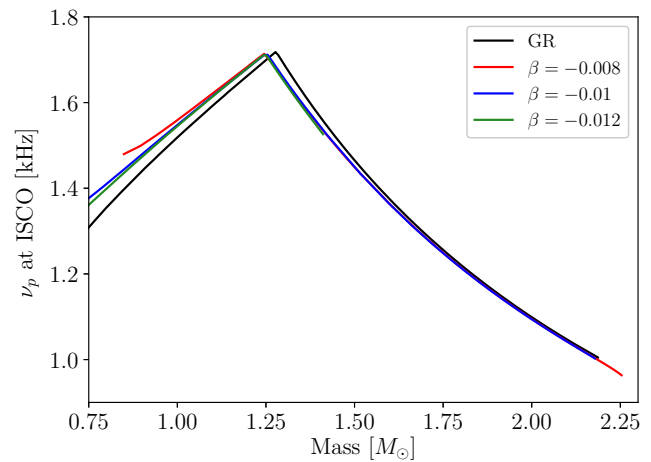}
	\caption{\label{Omp_Static_Plot} Orbital frequency at the ISCO radius as a function of the gravitational mass for GR and TMST with the monotonic coupling function (\ref{monotonic_coupling}), for $n=1$, different $\beta$ values and $f = 0$ Hz.}
\end{figure}

More important than the actual radius of the ISCO from observational standpoint is the orbital frequency obtained at its radius. 
Fig. \ref{Omp_Static_Plot} presents the results for the three cases of $\beta$ studied following the computation of eq. (\ref{Orbital_Omega}) in the Jordan frame of the theory at $f = 0$ Hz. Once again, deviations from GR are minor.
For masses exceeding $ 1.3 \ M_{\odot} $, the dependence for the topological neutron stars is essentially the same as that for the GR case, while for lower masses, particularly around $1 \ M_{\odot}$, the frequencies can be up to 3 \% higher than those predicted in GR. These results are not surprising since the deviation is in the region where ISCO is at the star's surface and Fig. \ref{MR_Relation_2} clearly indicates that the topological branch of neutron stars are more compact than their GR counterparts particularly at lower masses. This small difference, however, does not appear to be experimentally significant.

\begin{figure}[h!]
\centering
\includegraphics[scale=0.55]{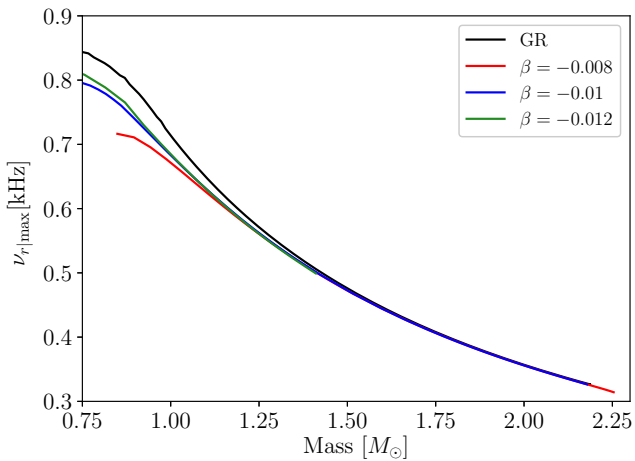}
\caption{\label{Omr_Static_Plot} Maximum absolute value of the radial epicyclic frequency $\nu_{r}$ as a function of the gravitational mass for GR and TMST with the monotonic coupling function (\ref{monotonic_coupling}), for $n=1$, different $\beta$ values and $f = 0$ Hz.}
\end{figure}

The radial epicyclic frequency's maximum value outside of the star at $f = 0$ Hz is given in the following Fig. \ref{Omr_Static_Plot}. The highest deviation is once again around $ 1 \ M_{\odot}$ reaching as much as 6\% and once again becoming negligible at higher masses. Since neutron stars with masses lower than that are not observed and currently considered as non-existing, the radial epicyclic frequency does not offer any significant observational difference between GR and topological neutron stars similarly to the orbital one.

\begin{figure}[h!]
\centering
\includegraphics[scale=0.55]{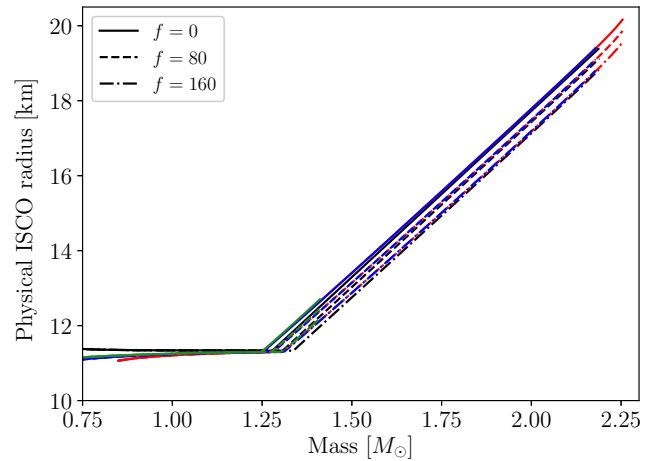}
\caption{\label{ISCO_Rotation_Plot} ISCO radius as a function of the gravitational mass for GR and TMST with the monotonic coupling function (\ref{monotonic_coupling}), for $n=1$ and different $\beta$ values. Solid lines denote static configuration while dashed and dash-dotted lines correspond to the $f=80$ Hz and $f=160$ Hz cases. Color scheme follows Fig.\ref{ISCO_Static_Plot}. }
\end{figure}

\begin{figure}[h!]
\centering
\includegraphics[scale=0.55]{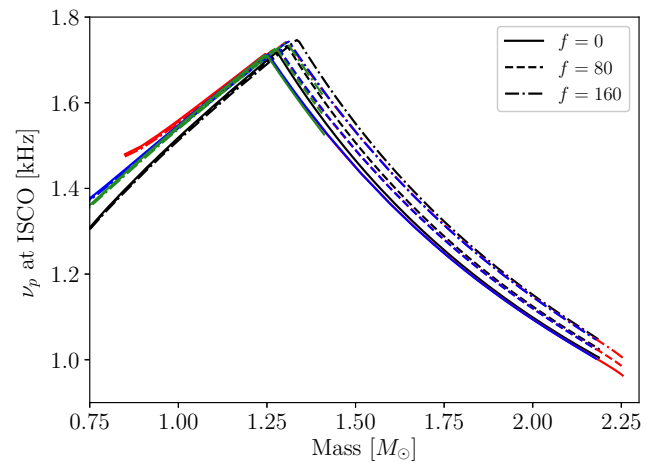}
\caption{\label{Omp_Rotation_Plot} Orbital frequency at the ISCO radius as a function of the gravitational mass for GR and TMST with the monotonic coupling function (\ref{monotonic_coupling}), for $n=1$ and different $\beta$ values. Solid lines denote static configuration while dashed and dash-dotted lines correspond to the $f=80$ Hz and $f=160$ Hz cases. Colors scheme follows Fig.\ref{Omp_Static_Plot}.}
\end{figure}

\begin{figure}[h!]
\centering
\includegraphics[scale=0.55]{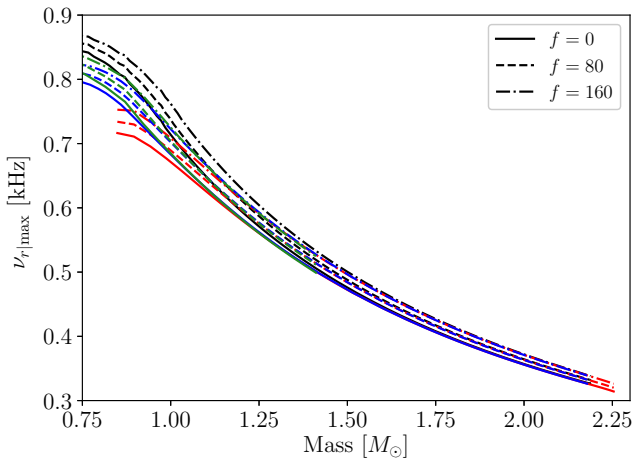}
\caption{\label{Omr_Rotation_Plot} Maximum absolute value of the radial epicyclic frequency $\nu_{r}$ as a function of the gravitational mass for GR and TMST with the monotonic coupling function (\ref{monotonic_coupling}), for $n=1$ and different $\beta$ values. Solid lines denote static configuration while dashed and dash-dotted lines correspond to the $f=80$ Hz and $f=160$ Hz cases. Colors scheme follows Fig. \ref{Omr_Static_Plot}.}
\end{figure}

The following three figures display the same quantities in the case of slow rotation.
The color schemes are the same as on figures \ref{ISCO_Static_Plot}, \ref{Omp_Static_Plot} and \ref{Omr_Static_Plot} with the solid lines corresponding to the $f=0$ Hz case, while the dashed and dash-dotted lines give the same dependencies at $f=80$ Hz and $f=160$ Hz.
The results are obtained in the slow rotation regime but the majority of observed neutron stars fall within this approximation.
There is also reason to believe that results at higher order of $\Omega$ are not qualitatively different \cite{Kalin_2_Uni}.
The ISCO for prograde orbit in all cases is shown on Fig.\ref{ISCO_Rotation_Plot}, followed by the orbital frequency and the radial epicyclic frequency on Figs. \ref{Omp_Rotation_Plot} and \ref{Omr_Rotation_Plot}.

The decrease in ISCO radius due to the star's rotation leads to an overlap between the plots in an even larger region making the GR and the topological cases harder to discern without knowing the rotational frequency precisely.
For each of the rotational cases by itself, no significant distinction is possible once again.

As far as the orbital and maximum radial epicyclic frequencies are concerned -- no major difference can be seen between GR and the $n=1$  topological neutron star branch with coupling function (\ref{monotonic_coupling}) for masses higher than $ \cong 1.2-1.3 M_{\odot}$.
Once again, in the region $~1-1.25 M_{\odot}$ minor quantitative differences on the order of 6 \% and less can be observed for both but these do not appear to be significant experimentally.

It appears that the monotonic coupling function (\ref{monotonic_coupling}) does not predict significant qualitative or quantitative deviations from GR as far as the X-ray observations are concerned. 
As discussed, other coupling functions, such as the non-monotonic one given by eq. (\ref{nmonotonic_coupling}), predict both qualitatively and quantitatively significantly different results from GR. 
The new effects that appear poses considerable interest and so it will be investigated further in a follow-up work.

\section{Conclusions}
In the present work we have further explored the properties of topological neutron stars (TNS) that is a new and very interesting class of compact objects in the Tensor Multi-Scalar Theories (TMST) of gravity. We considered the case of slow rotation and in addition we expanded the study of these objects' dependence on theory parameters, topological charge and coupling functions. The main focus of the paper was on the investigation of their moment of inertia and the quantities related to accretion under slow rotation.

We have confirmed the EOS independence between the suitably normalized moment of inertia for the TNS and their compactness, comparing the obtained results with those for neutron stars in General Relativity. It turns out that for one of the coupling functions, showing a monotonic behavior with respect to the scalar field, these universal relations are almost indistinguishable from GR, making them not only EOS independent, bus also theory of gravity independent up to a large extend. For more complicated non-monotonic functions, though, the deviations can be signification allowing not only to distinguish between GR and TMST, but also between solutions with different values of the topological charge. This demonstrates that the future observations can help us  test the very interesting hypothesis that neutron stars can posses a new property that is the topological charge.

We have also obtained accretion-relevant properties such as the ISCO radius, orbital and epicyclic frequencies for TNS at different rotation rates, comparing them once again to the GR results. The obtained TNS quantities differ significantly from those in GR only for some coupling functions and theory parameter values which leaves several observational traits to be sought after and explored. In the case of monotonic coupling function there are no qualitative difference between GR and TMST and in addition, the qualitative deviations from GR are very small, practically unmeasurable.  Some interesting qualitative differences appear in particular for non-monotonic coupling function for the ISCO-related quantities, where the quantitative deviations can be large as well, that will be further explored in a dedicated work as they require more profound investigations.
This provides a firm reason to pursue more accurate observations and systematic studies in order to constraint alternative theory parameters in the strong regime of gravity.

Further study of the higher order corrections to rotational neutron stars will additionally allow us to get their quadrupole moment and explore potential I-love-Q relations \cite{Yagi_1,Yagi_2} also opening new prospects of experimental confirmation through the gravitational waves observation data which is promising to become more abundant in the near-future \cite{LIGO_1,LIGO_2} 

\begin{acknowledgments}
DD acknowledges financial support via an Emmy Noether Research Group funded by the German Research Foundation (DFG) under grant
no. DO 1771/1-1. DD is indebted to the Baden-Wurttemberg Stiftung for the financial support of this research project by the Eliteprogramme for Postdocs. SY would like to thank the University of T\"ubingen for the financial support. The partial support by the Bulgarian NSF Grant
DCOST 01/6 and the networking support by the COST actions CA16104 and CA16214 is gratefully acknowledged.

\end{acknowledgments}

\newpage

\bibliography{apssamp}

\end{document}